\documentclass[aps,prl,reprint,showpacs,superscriptaddress,longbibliography]{revtex4-1}
\usepackage{mathtools,amssymb,graphicx,units}

\usepackage[plainpages=false,pdfpagelabels,colorlinks=true,linkcolor=red,urlcolor=blue,citecolor=blue,pdftitle={Title},pdfauthor={},pdfdisplaydoctitle=true,pdfduplex=DuplexFlipLongEdge]{hyperref}
\usepackage{verbatim}
\usepackage[english]{babel}
\usepackage{color}
\usepackage{ulem}

\newcommand{\beginsupplement}{%
        \setcounter{table}{0}
        \renewcommand{\thetable}{S\arabic{table}}%
        \setcounter{figure}{0}
        \renewcommand{\thefigure}{S\arabic{figure}}%
}

\graphicspath{{./fig/}}

\begin{document}
\title{Characterization of many-body mobility edges with random matrices}
\author{Xingbo Wei}
\affiliation{Zhejiang University, Hangzhou 310027, China}
\affiliation{Beijing Computational Science Research Center, Beijing 100193, China}
\affiliation{Westlake University, Hangzhou 310024, China}
\affiliation{Department of Physics, Zhejiang Normal University, Jinhua 321004, China}
\author{Rubem Mondaini}
%\email{rmondaini@csrc.ac.cn}
\affiliation{Beijing Computational Science Research Center, Beijing 100193, China}
%\author{Wei Zhu}
%\affiliation{Westlake University, Hangzhou 310024, China}
\author{Gao Xianlong}
%\email{gaoxl@zjnu.edu.cn}
\affiliation{Department of Physics, Zhejiang Normal University, Jinhua 321004, China}

\begin{abstract}
  Whether the many-body mobility edges can exist in a one-dimensional interacting quantum system is a controversial problem,  mainly hampered by the limited system sizes amenable to numerical simulations. We investigate the transition from chaos to localization by constructing a combined random matrix, which has two extremes, one of Gaussian orthogonal ensemble and the other of Poisson statistics, drawn from different distributions. We find that by fixing a scaling parameter, the mobility edges can exist while increasing the matrix dimension $D\rightarrow\infty$, depending on the distribution of matrix elements of the diagonal uncorrelated matrix. By applying those results to a specific one-dimensional isolated quantum system of random diagonal elements, we confirm the existence of a many-body mobility edge, connecting it with results on the onset of level repulsion extracted from ensembles of mixed random matrices.
\end{abstract}

\maketitle

\paragraph{Introduction.---}
A single-particle mobility edge characterizing the separation in energy between extended and localized states occurs in a variety of non-interacting quantum models, as in one-dimensional ones with long-range hoppings~\cite{Biddle2010,Biddle2011}, certain incommensurate modulations of the potential~\cite{Soukoulis1982,DasSarma1988,DasSarma1990,Ganeshan2015,Li2017}, or in three-dimensional disordered systems at a finite disorder strength~\cite{Bulka1985,Plyushchay2003}. The fate of mobility edges in the presence of interactions is of great interest to experimentalists and theorists, but as of today, a large debate questions the existence of its \textit{many-body} analogue. A variety of numerical results show the manifestation of many-body mobility edges in finite sizes, and some studies employing scaling analysis also argue that they should occur when approaching the thermodynamic limit~\cite{Laumann2014,Luitz2015,Xiaopeng2015,Jonas2014,Modak2015,Baygan2015,Bar2015,Serbyn2015,Mondragon2015,Nag2017,Xingbo2019}. Roeck \textit{et al.}, for example, contend this view, asserting that localized states will be eventually thermalized by ergodic ones, ruling out their concomitant appearance and consequently the occurrence of many-body mobility edges~\cite{Roeck2016,Roeck2017}. Thus, due to the size limitations used in all numerical calculations, consensus has not been achieved.

To provide a new angle in this controversy, we will step back, and rather than following the standard procedure of analyzing a typical physical model, we will head towards the theory of random matrices, which provides the basis of our understanding of quantum chaotic behavior and its eventual absence~\cite{Haake_book,Alessio2016}. These are naturally characterized by the rigidity of the matrix spectrum, with the former displaying a characteristic level repulsion that is not present in regular systems, which in turn, exhibit completely uncorrelated eigenvalues. When defining a gap between adjacent eigenvalues in the spectrum $S_j \equiv E_{j+1}-E_j$, chaotic and non-chaotic behaviors can be characterized by probability densities of the gaps $P_{\rm GOE}(S) = (\pi S/2)\exp{(-S^2\pi/4)}$ and $P_{\rm P}(S)=e^{-S}$, respectively, for unity mean spacing, when dealing with symmetric matrices~\cite{Porter_book}. Often what is done in studying the many-body localization (MBL) transition~\cite{Nandkishore2015,Altman2018,Abanin2019,Alet2018} for physical systems is to combine both types of matrices, gradually changing their relative weight until the statistics of the spectrum suddenly changes~\footnote{There have been discussions of whether this is ever possible, with some arguing that ergodic behavior survives at any finite value of disorder, i.e., at any finite weight of the Poisson matrix~\cite{suntajs2019}, while others claim that finite size effects are yet large to draw definitive conclusions~\cite{Abanin2019,Rajat2019,Sierant2019}.}. If increasing the weight of the matrix displaying Poisson statistics, this signals the onset of non-ergodic behavior once the eigenvalues of the combined matrix become uncorrelated, and their eigenfunctions display support not scaling with the matrix dimension~\cite{Alessio2016}. When purely dealing with random matrices, this type of combined ensemble mixing matrices of different symmetry classes has been studied in the past~\cite{Lenz1990,Lenz1991,Lenz1991b,Kota1999,Schierenberg2012,chavda2014poisson}, with corresponding surmises for the level spacing distributions being specifically defined in terms of the weights, and interpolating between $P_{\rm GOE}(S)$ and $P_{\rm P}(S)$, if, e.g., adding random matrices from the Gaussian orthogonal ensemble (GOE) and diagonal random matrices.

In this Letter, we employ the mixed ensemble model to study the tuning process from GOE to Poisson statistics, unlike others with intermediate ensemble models~\cite{Serbyn2016,Buijsman2019,Sierant2019b}, providing strong evidence on the existence of the mobility edges and the stability of the coexistence of localized and delocalized states in approaching the thermodynamic limit.

\begin{figure}[htbp]
  \includegraphics[width=1\columnwidth]{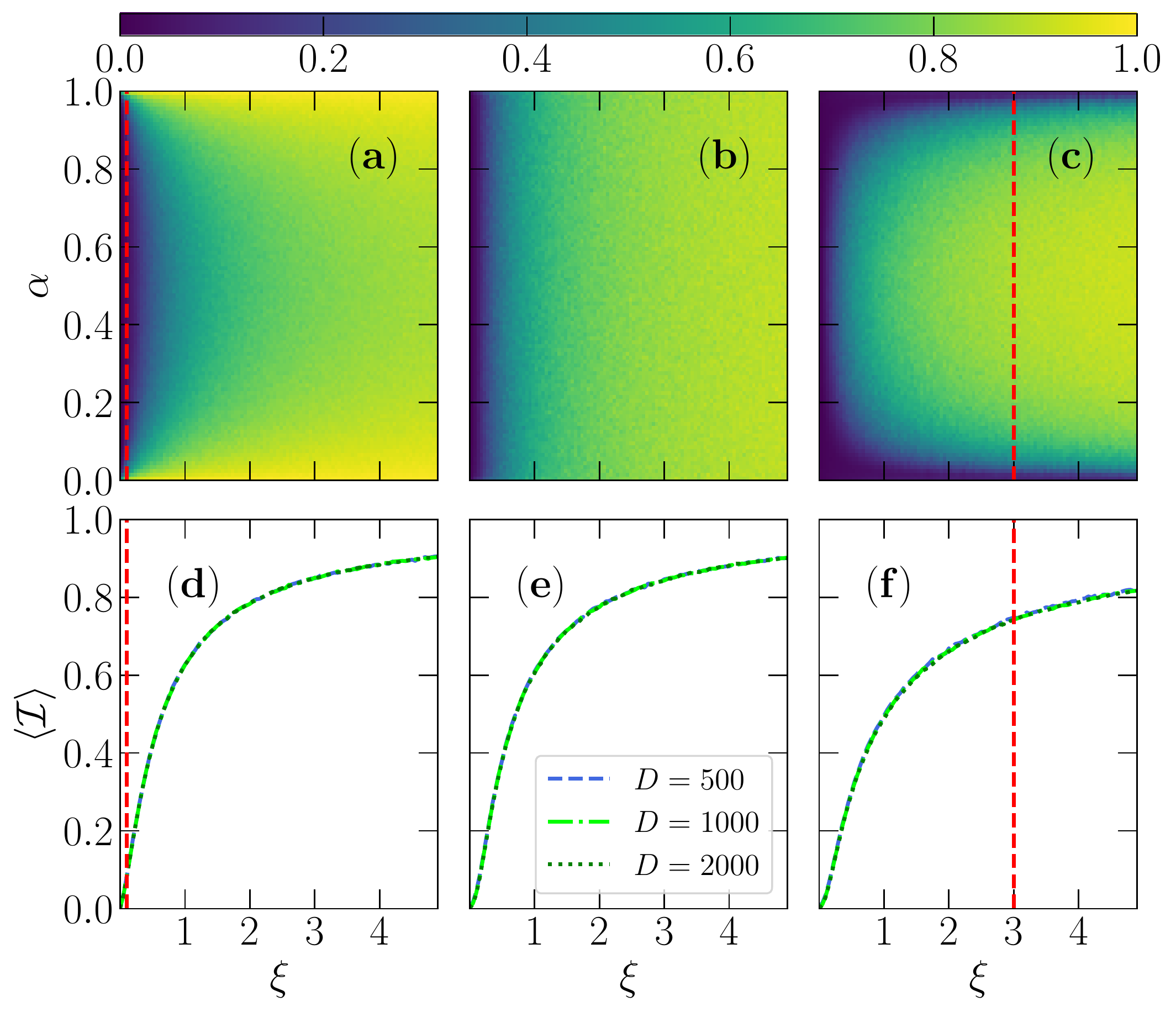}
 \vspace{-0.6cm}
 \caption{Phase diagrams and corresponding average (over the whole spectrum) inverse participation ratio $\langle {\cal I} \rangle$  as a function of $\xi$ for different ${\sf M}_i$s. The diagonal elements of ${\sf M}_i$ are drawn from a  Gaussian distribution in (a) and (d); uniform distribution in (b) and (e); cosine function $F(x) = \sqrt{2}\cos(2\pi x) $ with a random number $x\in[0,1]$ in (c) and (f). Color bar represents the value of IPR. Disorder realizations have been performed. The ordinate $\alpha= i/(D-1)$ is the normalized level position in (a), (b) and (c) with $i=0, 1, ..., D-1$ the energy level and the ordinate window $\Delta{\alpha}=0.01$. Red dashed lines mark the positions of $\xi$ ($\xi=0.1$ in (a) and (d); $\xi=3$ in (c) and (f)), further analyzed in Fig. \ref{fig:2}.}
 \label{fig:1}
\end{figure}

\paragraph{Model.---}We construct a combined random matrix ${\sf M}$, which reads
\begin{eqnarray}
\label{e1}
{\sf M}=(1-k){\sf M}_e+k{\sf M}_i,
\end{eqnarray}
where ${\sf M}_e$ is a GOE random matrix constructed by means of a normal random distribution matrix ${\sf A}$ by $({\sf A}+{\sf A}^T)/2$, ${\sf M}_i$ is a diagonal matrix with random elements, and $k\in[0,1]$ is a real coefficient. The limiting values $k = 0$ (${\sf M}={\sf M}_e$) and $1$ (${\sf M}={\sf M}_i$) characterize chaotic and non-chaotic regimes, where the  probability distributions of gaps in the spectrum follow $P_{\rm GOE}$ and $P_{\rm P}$, respectively. Our main interest is focused on the intermediate case $0<k<1$, where ${\sf M}$ can potentially describe the transition from chaos to localization. In this regime of mixed ensembles, surmises for the probability distribution have been derived~\cite{Lenz1991,Lenz1991b,Schierenberg2012}, in direct similarity with the Wigner surmises of pure ensembles~\cite{Haake_book}. In what follows, we investigate two main indicators of chaotic matrices, as to classify both the eigenvalues spectrum as well as their eigenfunctions. For the first, we notice that the gap distributions albeit precise in characterizing the ergodic properties of the spectrum, suffer from technical difficulties related to the necessity of having mean level spacing $\langle S \rangle$ constrained to unity for the aforementioned distributions to be valid. Thus instead, we look at the distribution $P(r)$ of the ratio of two consecutive gaps in the spectrum $r_\alpha = \min\{S_\alpha,S_{\alpha+1}\}/\max\{S_\alpha,S_{\alpha+1}\}$, requiring no unfolding schemes~ \cite{Atas2013}.

In turn, a direct measure of the (de)localization of the eigenfunctions is obtained by the inverse participation ratio (IPR), ${\cal I}^{(\alpha)}=\sum_n|\psi_n^{(\alpha)}|^4$, where $\psi_n^{(\alpha)}$ is the $\alpha$-th eigenstate of the matrix and $n$ is the basis state index. The eigenstate properties are characterized according to its limiting values as
\begin{equation}
  \lim_{D\rightarrow\infty}{\cal I}^{(\alpha)} \propto \left\{
\begin{aligned}
    1/D,  \qquad   &\rm{chaotic \, (extended)}  \\
    \rm{const}., \qquad   &\rm{localized} \\
\end{aligned}
\right.
\end{equation}
where $D$ is the matrix dimension. The IPR of a chaotic (extended) eigenstate is $D$-dependent, in contrast to a localized one; therefore, a finite-size scaling is necessary to differentiate the two types of states when approaching the limit $D\rightarrow\infty$. In that regime, it is important to extract a scaling parameter that can account for the modifications the mixed ensemble suffers. For example, the spectral variance of ${\sf M}_e$ is approximately proportional to $D$ for large enough matrix dimension~\cite{chavda2014poisson}. For that, we conjecture the following scaling parameter~\footnote{See \cite{chavda2014poisson} for a similar conjecture inspired by the preliminary derivation of probability distribution of eigenvalues of the mixed ensembles.},
\begin{eqnarray}
 \label{xixi}
 \xi=\frac{k\tilde{\sigma}}{(1-k)D\sigma},
\end{eqnarray}
where $\tilde{\sigma}$ and $\sigma$ are the standard deviation of the diagonal elements in ${\sf M}_i$ and ${\sf M}_e$, respectively. When $k$ changes from $0$ to $1$, $\xi$ changes from $0$ to infinity. Without loss of generality, we set $\tilde{\sigma}=\sigma=1$ and choose the standard deviation of non-diagonal elements $\sqrt{2}/2$ times of the diagonal elements in ${\sf M}_e$. We find that, by changing $k$ and $D$, but fixing $\xi$, the system keeps the same local and chaotic properties.

\begin{figure}[t]
  \includegraphics[width=1\columnwidth]{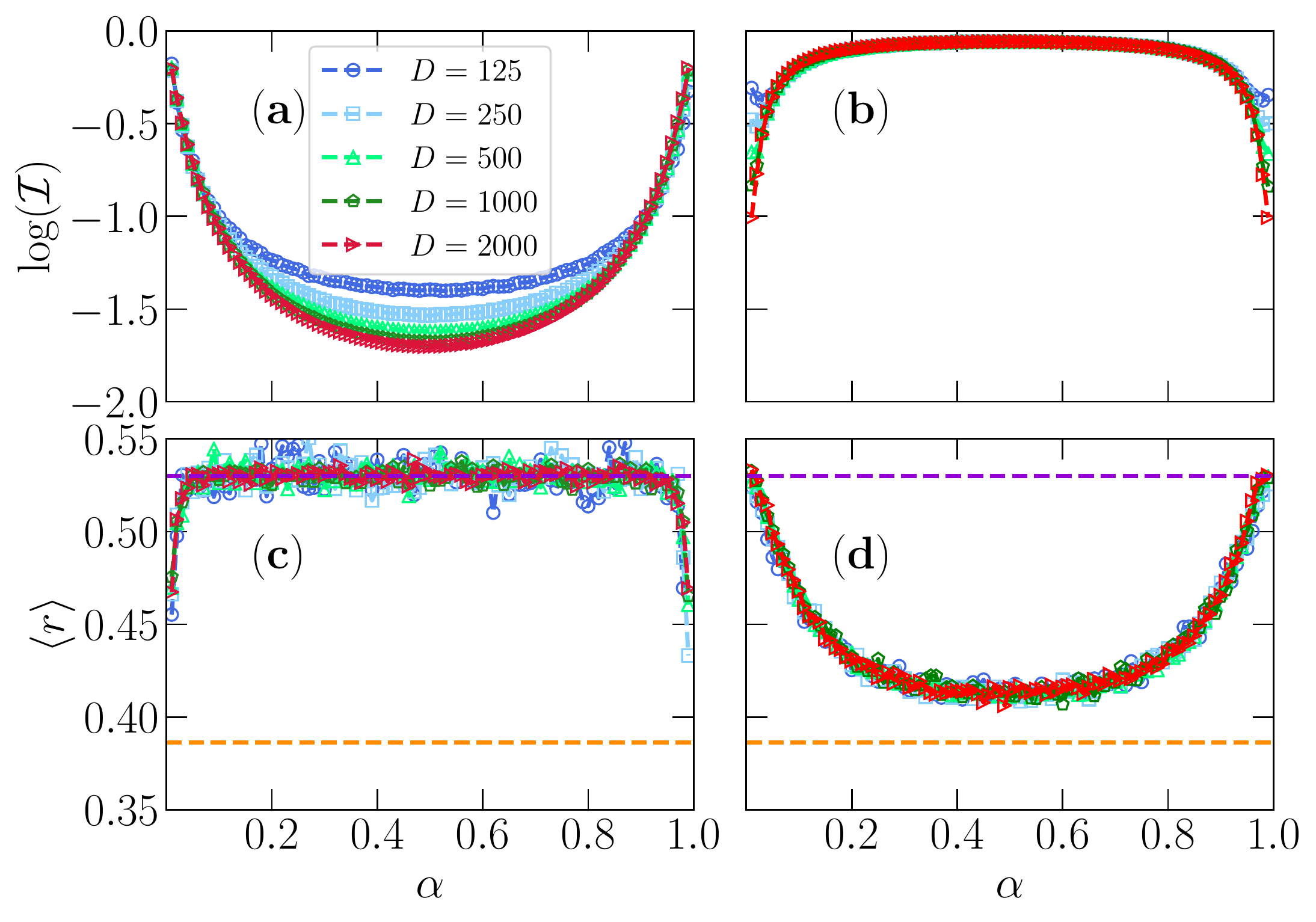}
 \vspace{-0.6cm}
 \caption{Inverse participation ratio ${\cal I}^{(\alpha)}$ and corresponding average ratio of adjacent energy gaps $\langle r \rangle$ as a function of $\alpha$ for different $D$s. $\xi=0.1$ for (a) and (c) with ${\sf M}_i$ of Gaussian distribution; $\xi=3$ for (b) and (d) with with ${\sf M}_i$ from the cosine distribution function. The purple dashed line marks $\langle r \rangle = 0.53$ (GOE) whereas the  orange dashed one marks $\langle r \rangle =2 \ln 2-1$ (Poisson) in (c) and (d), respectively.}
 \label{fig:2}
\end{figure}

We test this scaling form by employing specific distributions for the diagonal elements in ${\sf M}_i$, Gaussian, uniform and cosine distributions in Fig.~\ref{fig:1} [(a),(d)],~\ref{fig:1} [(b),(e)] and \ref{fig:1} [(c),(f)], respectively. As shown in Fig.~\ref{fig:1}(d), ~\ref{fig:1}(e), ~\ref{fig:1}(f), the average inverse participation ratios over the whole spectrum, $\langle \cal I \rangle$, are completely coincident for $D=500$, $1000$ and $2000$ when using the scaling parameter $\xi$. This is indicative that for large matrices, the ${\cal I}^{(\alpha)}$ diagram for different distributions  Fig.~\ref{fig:1}(a), ~\ref{fig:1}(b), ~\ref{fig:1}(c) (for $D=2000$) correctly reflect the phase diagrams for $D\rightarrow\infty$ (See Fig.~\ref{fig:2} for details of IPR before the average for the different $D$s). Interestingly, the chaotic region shows a ``D''-like shape in Fig.~\ref{fig:1}(a), and a ``C"-like shape in Fig.~\ref{fig:1}(c). Since out of the closure of ``C'' and ``D'' it is a localized phase, both situations indicate a typical phase diagram of mobility edges. However, there is no obvious mobility edges for the uniformly distributed ${\sf M}_i$ [Fig.~\ref{fig:1}(b)]. Thus, those are model-dependent and do exist for ${\sf M}_i$ of Gaussian distribution or of cosine distribution for fixed $\xi$ even when the dimension of the matrix goes to infinity. Moreover, the transition from classical chaos to localization with increasing $\xi$ has characteristics of a crossover.

\begin{figure}[t]
  \includegraphics[width=1\columnwidth]{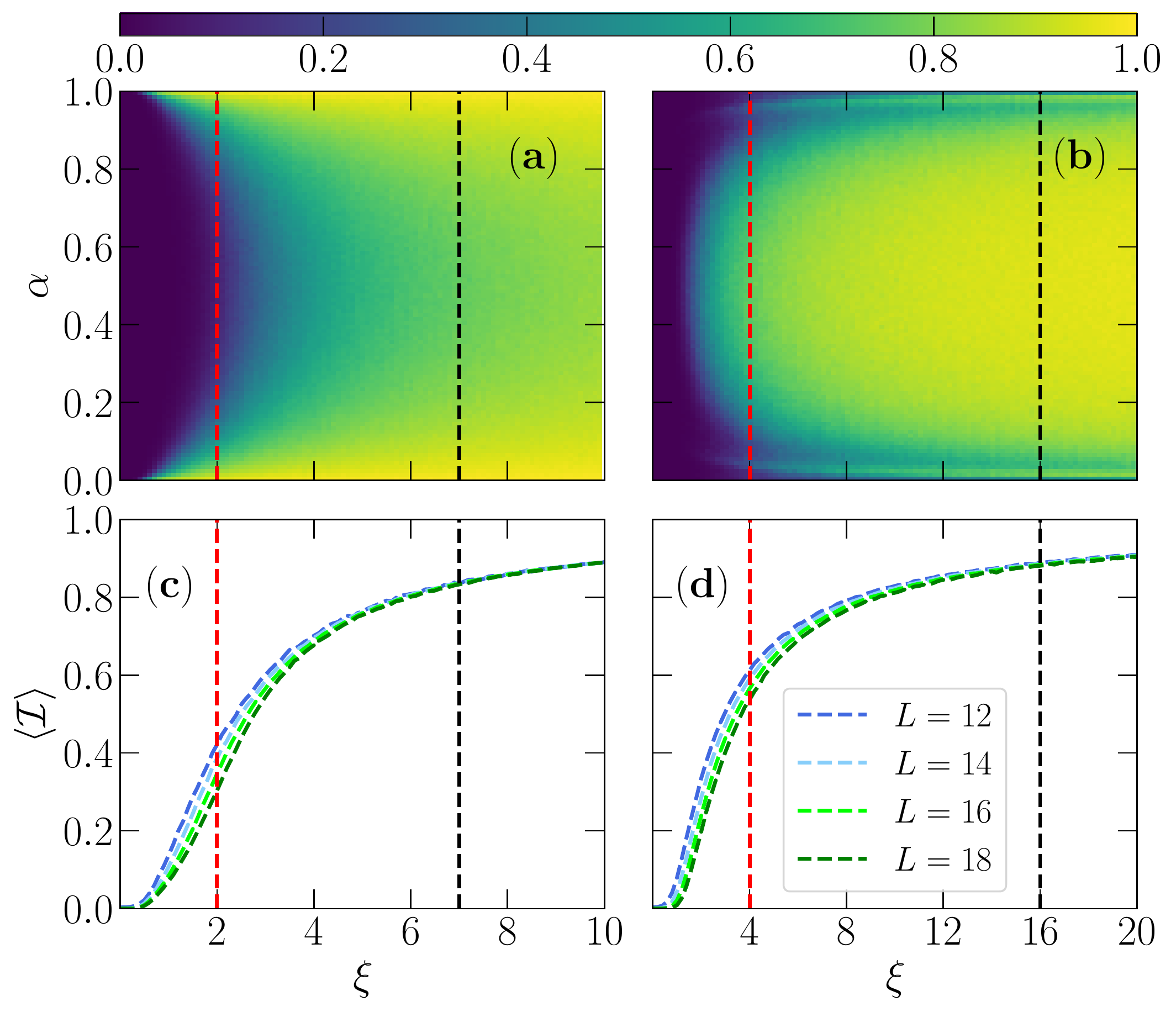}
 \vspace{-0.6cm}
 \caption{The ${\cal I}^{(\alpha)}$ phase diagrams for $L=18$ and $\langle {\cal I} \rangle$ as a function of $\xi$ for different $L$s at half filling. (a) and (c) for ${\sf G}$ of Gaussian distribution; (b) and (d) for ${\sf G}$ of cosine function distribution, both with $\tilde{\sigma}=1.0$. Color bar represents the value of ${\cal I}$. The ordinate window $\Delta{\alpha}=0.01$ in (a) and (b). Red and black dashed lines mark the positions of $\xi$ ($\xi=2.0, 7.0$ in (a) and (c), $\xi=4.0, 16.0$ in (b) and (d), respectively. See further analysis in Fig. \ref{fig:4}.)}
 \label{fig:3}
\end{figure}

We further study the IPR and $\langle r \rangle$ with fixed $\xi$ but different $D$s in Fig.~\ref{fig:2}. The latter, apart from finite size fluctuations, is coincident for increasing dimensions $D$. For matrices ${\sf M}_i$ from the Gaussian distribution, the states at the edges of the spectrum ($\alpha\lesssim0.05$ and $\alpha\gtrsim0.95$, with $\xi = 0.1$) do not follow GOE statistics, indicating that these  are not chaotic. This is supported by the corresponding values of the IPR with increasing matrix sizes in Fig.~\ref{fig:2}(a): Only at the central part of the spectrum scaling with $D$ occurs. Conversely, the situation is opposite for ${\sf M}_i$ constructed from a cosine function distribution [See Figs.~\ref{fig:2}(b) and \ref{fig:2}(d)]. At the edges of the spectrum, ${\cal I}^{(\alpha)}$ decreases with increasing dimensions and is accompanied by a mean ratio of adjacent gaps characteristic of chaotic matrics from the GOE ensemble for the value of $\xi=3$ used. Based on these results for mixed ensembles of random matrices, we apply similar formalism to Hamiltonian matrices of typical quantum systems.

\noindent\noindent\paragraph{Many-body mobility edges.---} In order to test our analysis and to determine whether a many-body mobility edge can exist in a one-dimensional quantum system on a lattice, we construct a many-body model, composed of $N$ spinless fermions in a chain with $L$ sites. The Hamiltonian is given by
\begin{eqnarray}
 \label{eq:hamilt}
  \hat{H}=\hat{T}+\hat{V},
\end{eqnarray}
where $\hat{T}$ is the (non-random) nearest-neighbor hopping matrix,
\begin{eqnarray}
 \hat T = -&t&\sum_{i=1}^L \left(\hat c_i^\dagger \hat c^{\phantom{}}_{i+1}+ {\rm H.c.}\right),
\end{eqnarray}
assuming $\hat c_{L+1} = \hat c_1$. The spectral variance of $\hat T$ is proportional to $N$ for large matrix sizes $D = {L\choose N}$ and fixed filling $n=N/L$, as $\sigma_E^2=2N(1-n)$, which contrasts with a typical GOE matrix where $\sigma_E^ 2\propto D$. In turn, the second matrix is diagonal, and characterizes both the interaction and disorder, which is typically the case for Fock bases in real space. It assumes a form $\hat{V}=V {\sf G}$, with $V$ its strength, and ${\sf G}$ populated with random numbers selected from a particular distribution. Motivated by the cases where mobility edges were found when dealing with pure random matrices, we focus on two types: Gaussian and cosine distributions, both characterized by a standard deviation $\tilde{\sigma}$ of its elements. Other distributions, as power-law, can also result in mobility edges and are analyzed in the Supplemental Material~\cite{SM}. We notice  that short-range homogeneous interactions result in diagonal matrices whose distribution of elements is given by a discrete Gaussian, and in the case the interactions are fully random, they approach a continuous distribution that can also lead to the onset of MBL~\cite{Sierant2017, Li2017b}. Furthermore, a typical Anderson-like disorder with uniform random local energies also result in a Gaussian distribution of entries in the diagonal elements for a many-body basis.

The scaling parameter of Eq.~\eqref{xixi} in this model becomes,
\begin{eqnarray}
 \label{eq:xi}
 \xi=\frac{V\tilde{\sigma}}{tN},
\end{eqnarray}
which we use, as before, to scale the average values of IPR for increasing matrix sizes. Hereafter, the filling number is set to $n=N/L=0.5$, which results in the largest possible matrix sizes for a given $L$.

\begin{figure}[t]
  \includegraphics[width=1\columnwidth]{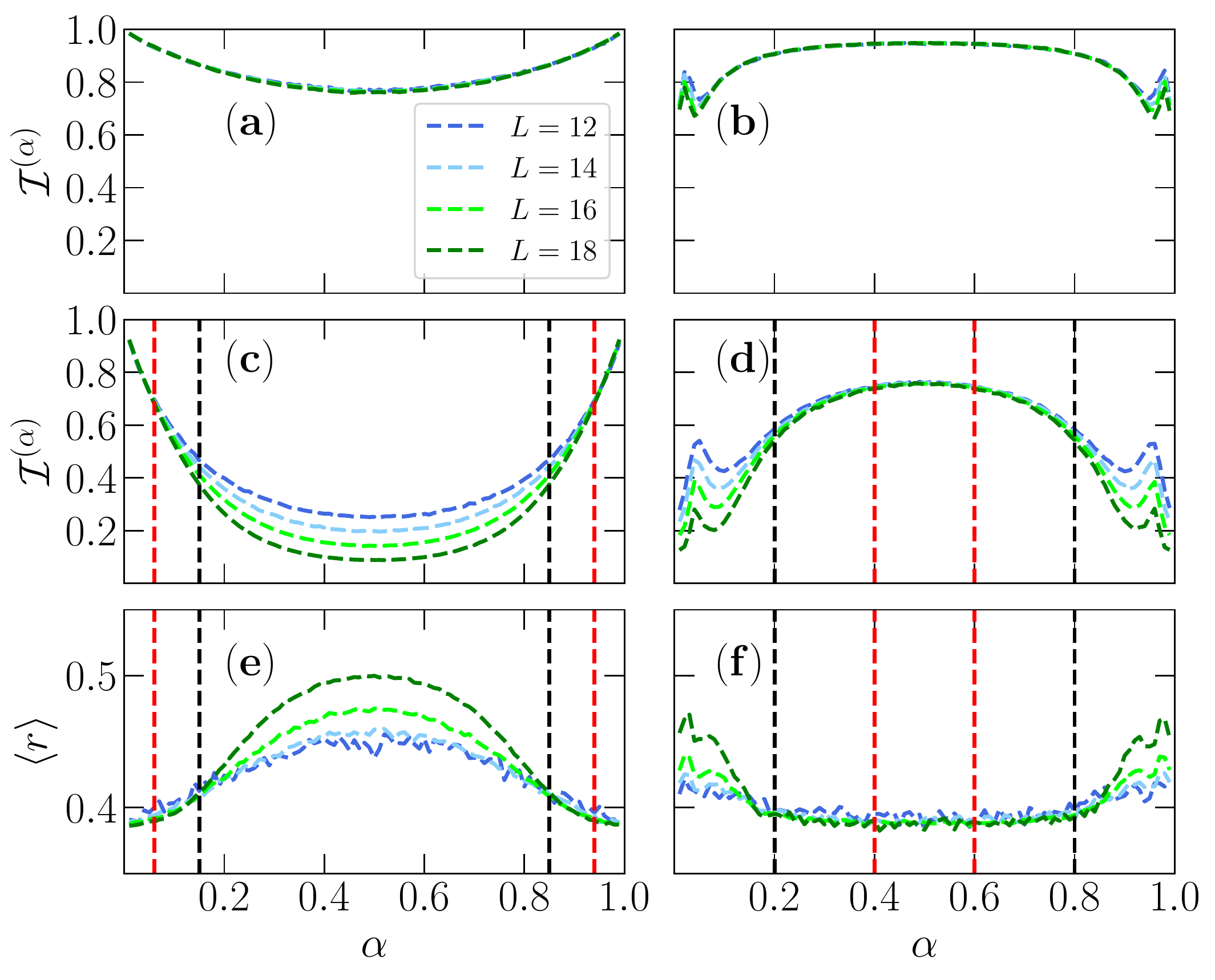}
 \vspace{-0.6cm}
 \caption{ IPR  and $\langle r \rangle$ as a function of $\alpha$ for different $L$s. (a) and (c) correspond to Fig.~\ref{fig:3}(c) with $\xi=7.0$ and $\xi=2.0$, respectively. (b) and (d) correspond to Fig.~\ref{fig:3}(d) with $\xi=16.0$ and $\xi=4.0$, respectively. (e) and (f) are the corresponding $\langle r \rangle$ value of (c) and (d), respectively. The red dashed lines mark IPR separation points and the black dashed lines mark $\langle r \rangle$ separation points in (c), (d), (e), and (f). In (c) and (e), the red dashed lines are at $\alpha=0.06$ and $\alpha=0.94$, and the black dashed lines $\alpha=0.15$ and $\alpha=0.85$. In (d) and (f), the red dashed lines are at $\alpha=0.40$ and $\alpha=0.60$ and black dotted lines $\alpha=0.20$ and $\alpha=0.80$. }
 \label{fig:4}
\end{figure}

In Fig.~\ref{fig:3}, we calculate the phase diagrams of the IPR for $L=18$ and its average value, $\langle {\cal I} \rangle$, as a function of $\xi$ for different $L$s. The phase diagrams in Fig.~\ref{fig:3}(a), ~\ref{fig:3}(b) are similar to those of Fig.~\ref{fig:1}(a), ~\ref{fig:1}(c) for the pure random matrices, showing robust many-body mobility edges for Gaussian and cosine function distributions of diagonal matrix elements, respectively. In Fig.~\ref{fig:3}(c), $\langle {\cal I} \rangle$ shows converged values for different system (matrix) sizes for $\xi\gtrsim 7$ [details shown in Fig.~\ref{fig:4}(a)], indicating Eq.~\eqref{eq:xi} serves as a scaling parameter, and that the system size $L$ here is no longer an important factor to trigger MBL. For $\xi < 7$, $\langle {\cal I} \rangle$ decreases as $L$ increases, a signal that ergodic states start to populate the spectrum. Figure \ref{fig:3} (d) for the cosine distribution shows very similar behavior, but with a critical $\xi \simeq 16$ separating the regime with the presence of ergodicity and full MBL.

In Fig.~\ref{fig:4}, we investigate the system size effects on ${\cal I}^{(\alpha)}$ and $\langle r \rangle$ across the spectrum $\alpha$, for different typical values of $\xi$ in both distributions. For the many-body localized region [typical value $\xi = 7$ (16) for Gaussian (cosine) distributions], Fig.~\ref{fig:4} (a) and (b) show that completely coincident IPR for different sizes confirm localization across the whole spectrum. For smaller weights $\xi$ of the diagonal matrices, ergodic and delocalized states naturally occur in the central part (edges) of the spectrum for ${\sf G}$ of the Gaussian (cosine function) distribution. These regions are identified by an $L$-dependence in both ${\cal I}^{(\alpha)}$ and $\langle r \rangle$, with the former decreasing with the system size and the latter heading towards $\langle r \rangle_{\rm GOE}$ for increasing $L$s. When these quantities display system size independence, lack of ergodicity and localization ensues. We notice, however, that the onset of this behavior is not coincident in $\alpha$, i.e., it suggests that the many-body mobility edge occurs not as a sharp transition in the spectrum but rather as one characterized by an intermediate phase with non-ergodic but delocalized wavefunctions. This regime has been found in physical models and dubbed non-ergodic metal~\cite{Xiaopeng2015,Li_16,Deng2017,Hsu_2018}.

\begin{figure}[htbp]
  \includegraphics[width=1\columnwidth]{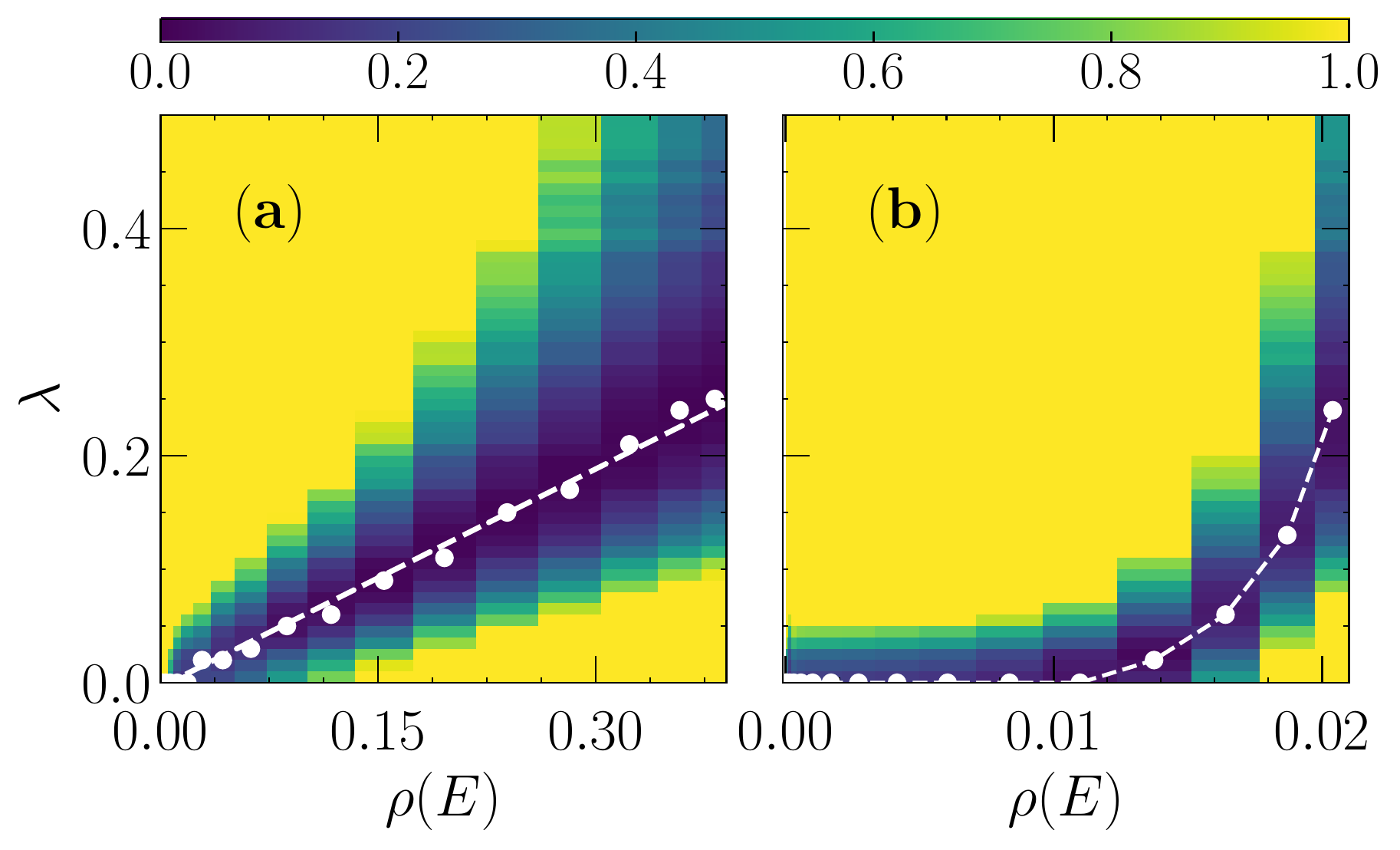}
 \vspace{-0.6cm}
 \caption{The distance $\chi$ between the numerical surmise of $3\times3$ random matrices and the density-of-states-resolved distribution of the ratio of adjacent gaps. In (a) for a combined matrix ${\sf M}_i + \Lambda{\sf M_e}$ ($\Lambda = 1/D$), whereas in (b) for the physical system $\hat H = \hat T + \hat V$, with $\xi = 2$ and $L =18$ -- different system sizes are attempted in~\cite{SM}.}
 \label{fig:5}
\end{figure}

\paragraph{Mobility edges and predictions of mixed ensembles.---}
Mixed ensembles provide further fundamental insights on the onset of ergodicity. Starting from a Poisson diagonal matrix, any non-zero weight of an added GOE matrix triggers a finite degree of level repulsion~\cite{Lenz1991,Schierenberg2012}, a defining characteristic of ergodicity. However, the chaotic effects induced by the latter crucially depend on the density of states. That is, the effective weight of the GOE matrix on the resulting mixed matrix grows as denser the spectrum is~\cite{Schierenberg2012}, initially investigated via the analysis of the density-of-states-resolved statistics of the gaps $P(\rho,S)$, which requires unfolding of the spectrum. An $r$-statistic analysis avoids this inconvenience, but a surmise for $P(r)$ in the mixed GOE-Poisson ensemble is yet elusive~\cite{chavda2014poisson}. Nonetheless, we can infer similar information regarding the effect of the GOE random matrix by a working numerical definition. Starting from random $3\times3$ matrices of the mixed ensemble form ${\sf M} = {\sf M}_i + \lambda {\sf M}_{e}$, one can obtain the \textit{numerical} surmise $P(r,\lambda)$ by using a set of 2,000 of such matrices for each $\lambda$. Subsequently, we extract the adjacent gap distribution for a much larger combined $D\times D$ matrix with a typical weight $\lambda = \Lambda$. By dividing the spectrum in a large number of homogeneous energy windows, we resolve the distribution by the density of states $\rho$, obtaining $P(r, \rho)$. Finally, the effect of the weight $\lambda$ of the chaotic matrix can be quantified as,
\begin{equation}
 \chi \equiv \sum_r [P_{3 \times 3}(r, \lambda) - P_{2000 \times 2000}(r, \rho) ]^2,
\end{equation}
which is shown in Fig.~\ref{fig:5}, as a color plot. The minima of this work definition (white markers) assures that the effect of the $\lambda$ weight of the GOE matrix is linearly proportional to the density of states, a result also valid for the distributions of the gaps $P(\rho,S)$~\cite{Schierenberg2012}. We now apply the same analysis in the case of a physical system [Eq.~\eqref{eq:hamilt}]. Now, the different density of states results in a remarkable contrasting behavior for the effect of the thermalizing matrix. $\lambda$ approaches zero at \textit{finite} density of states, which in practice results in the \textit{absence of ergodicity} at the edges of the spectrum, that are typically less dense. This thus indicates that a sharp transition occurs in the ergodic properties of the eigenvalues in the spectrum, where a region with small density of states does not suffer from thermalizing effects and can be associated to the manifestation of a mobility edge.

\paragraph{Summary and discussion.---} We study the process from GOE to Poisson statistics by constructing a combined random matrix as to shed light on the long lasting debate on the existence of mobility edges. We find phase diagrams in the thermodynamic, determined by the invariant scaling parameter $\xi$, suggesting its existence for different distributions of the diagonal random matrix. Lastly, by employing a numerical analysis on the effect of the strength of the GOE matrix in different parts of the spectrum, we infer that less dense regions have an effective vanishing weight of the ergodic matrix, albeit denser ones do render ergodic behavior. Thus, we argue that mobility edges can exist in physical systems with short-ranged hoppings, even when approaching the thermodynamic limit. A closed form of the probability distribution of the ratio of adjacent gaps of mixed ensembles may result in an even more systematic way to define whether different quantum systems may or not display mobility edges and will be reserved for a future work.

\begin{acknowledgments}

  The authors acknowledges insightful discussions with T.~Wettig, W. Zhu, and M. Gong. GX and WX acknowledge support from NSFC under
  Grants No. 11835011 and No. 11774316. RM acknowledges support from NSFC Grants No. U1930402, No. 11674021, 11851110757 and No. 11974039. The computations were performed in the Tianhe-2JK at the Beijing Computational Science Research Center (CSRC).
\end{acknowledgments}
\bibliography{RMT_MMEv5}

\clearpage

\onecolumngrid

\begin{center}

{\large \bf Supplementary Materials:
\\ Characterization of many-body mobility edges with random matrices}\\

\vspace{0.3cm}

\end{center}

\vspace{0.6cm}

\twocolumngrid

\beginsupplement

\paragraph{Many-body mobility edges for other distribution of the diagonal random matrix: Power-law function.---}
\begin{figure}[htbp]
  \includegraphics[width=1\columnwidth]{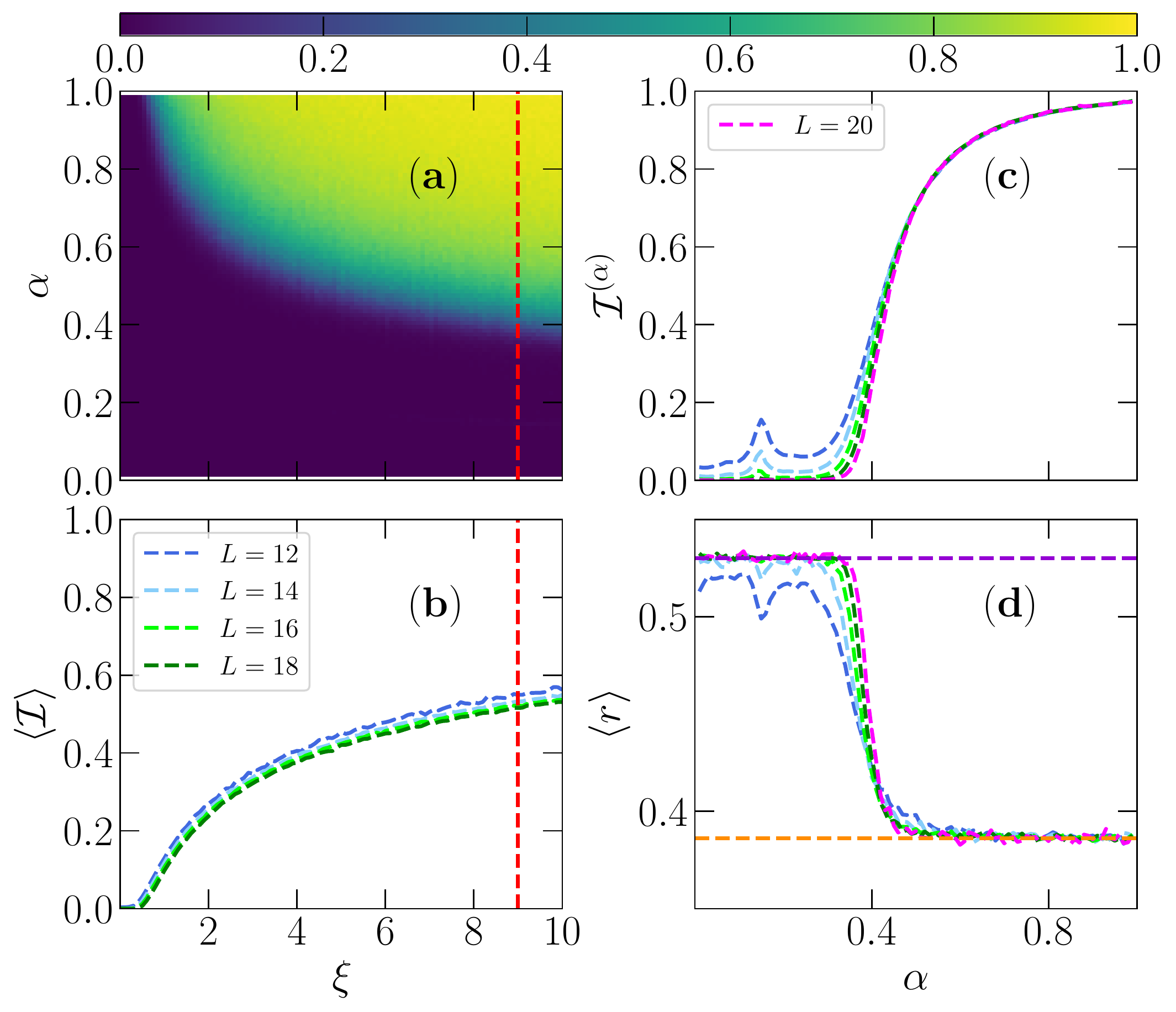}
 \vspace{-0.6cm}
 \caption{The phase diagram for $L=18$ (a) and average IPR $\langle {\cal I} \rangle$ (b) as a function of $\xi$ for different $L$'s with ${\sf G}$ populated with elements from a power-law function distribution. The red dashed lines mark $\xi = 9.0$, used in the remaining panels. Spectrum resolved IPR and $\langle r \rangle$ for $\xi = 9.0$ in (c) and (d), respectively.  The purple horizontal dashed line marks the GOE mean ratio of adjacent gaps, $\langle r \rangle =0.53$, and orange one the correspondent Poisson value $\langle r \rangle =2\ln2-1$ in (d).}
 \label{fig:5_old}
\end{figure}

In the main text, we have analyzed how the distribution of the random diagonal elements affect the overall shape of the localization-delocalization transition in the spectrum-disorder amplitude ($\alpha$ vs. $\xi$) space. In both cases where a mobility edge was observed, Gaussian and cosine function distribution, it is manifested mainly at the edges of the spectrum, where the density of states is small. Finite size effects are thus more dramatic at those regimes, as increasing $L$'s can significantly alter the sparsity of the levels at the ends of the spectrum. Since the ergodic-MBL transition shape significantly depends on the random distribution used to construct $\sf G$, we can proceed with the exercise of finding another distribution whose manifestation of a mobility edge avoids this problem. A natural choice, beyond the ones we have chosen, is a power-law function, which we select as $F(x) = (x^4-1/5)/(4/15)$. Here $x$ is a random number, picked in the interval $[-1,1]$, and $1/5$, and $4/15$ are the mean value and standard deviation of $F(x) = x^4$, respectively.

In Fig.~\ref{fig:5_old}(a), we report the phase diagram for the physical model \eqref{eq:hamilt}, with $L=18$, and elements in $G$ selected according to the probability distribution $F(x)$ above. There is a clear many-body mobility edge with only one ergodic-to-MBL transition, and the critical point is around the middle of the spectrum for $\xi=9$.  Fig.~\ref{fig:5_old}(b) indicates that the average IPR, $\langle {\cal I} \rangle$, possesses a small $L$-dependence in a large range of $\xi$, in similarity to the other distributions studied in Fig.~\ref{fig:3}. However, the spectrum resolved IPR, ${\cal I}^{(\alpha)}$, displays, on the other hand, a much more sharp transition in the spectrum approaching zero much faster than for other distributions studied in the main text. This sharp transition on the localization properties of the eigenfunctions is accompanied by a well marked transition on the level repulsion [Fig.~\ref{fig:5_old}(d)], displaying a GOE-to-Poisson transition for the statistics of the level spacings at similar values of $\alpha$.

\paragraph{Scaling parameter $\xi$ and relation to previous results.---}
In Ref.~\onlinecite{chavda2014poisson}, Chavda \textit{et al.} used a mixed ensemble of the form
\begin{equation}
{\sf H}_{\lambda}={\sf H}_0+\lambda {\sf V},
\end{equation}
where $\sf H_0$ is the Poisson matrix and $\sf V$ the corresponding GOE one. Similarly to what we did, they selected the variance of the off-diagonal elements of the GOE matrix being one-half of the variance of the diagonal ones, $\sigma^2$. After that, by arriving on a non-closed form of the joint probability of the eigenvalues of a 3$\times$3 ensemble, they notice that it depends on the transition parameter,
\begin{equation}
 \Lambda = \frac{\lambda^2\sigma^2}{2\overline{d_0}^2},
\end{equation}
where $\overline{d_0}$ is the mean level spacing between nearest eigenvalues of the diagonal (unperturbed) Poisson matrix. To match our definition of the mixed ensemble, one can easily identify that $\lambda = \frac{1-k}{k}$ and that the mean level spacing of the Poisson matrix $\overline{d_0} = \frac{\tilde \sigma}{\sqrt{2}D}$. With this, the scaling parameter we used and theirs are related as,
\begin{equation}
 \xi = \frac{1}{\sqrt{\Lambda}}.
\end{equation}

As us, they identify the very small dependence of the average ratio of adjacent gaps over the whole spectrum on the matrix size $D$, when scaling with the transition parameter $\Lambda$, and we do so by looking at the average value of the IPR, scaling it with $\xi$ (see Fig.~\ref{fig:1}).

\begin{figure}[htbp]
  \includegraphics[width=1\columnwidth]{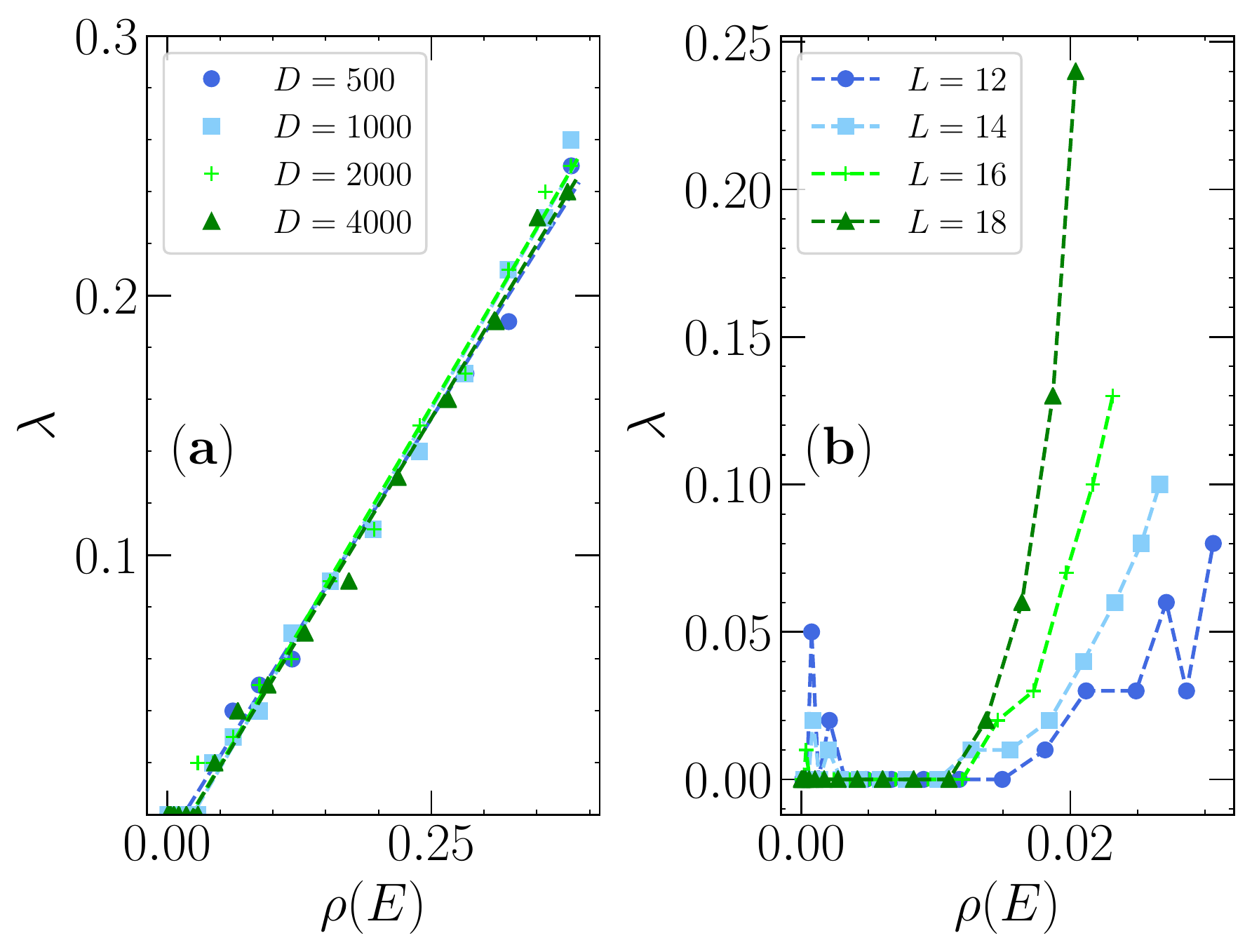}
 \vspace{-0.6cm}
 \caption{The effective weight of the ergodic matrix in the combined ensemble as a function of the normalized density of states: The goal is to analyze the effects of matrix finiteness. In (a), for the combined matrix
 ${\sf M}_i + \Lambda{\sf M_e}$ ($\Lambda = 1/D$), whereas in (b) for the physical system $\hat H = \hat T + \hat V$, with $\xi = 2$. As in the main text, the markers are obtained by the minimization of the distance $\chi$ between the numerical surmise in small matrices and the one obtained for the different matrix sizes, resolving by the density of states. The lines in panel (a) are linear fittings when neglecting the vanishingly small densities. In (b), large fluctuations can also occur in this regime, in special for the small matrix sizes.}
 \label{fig:S2}
\end{figure}

\begin{figure}[t!]
  \includegraphics[width=1\columnwidth]{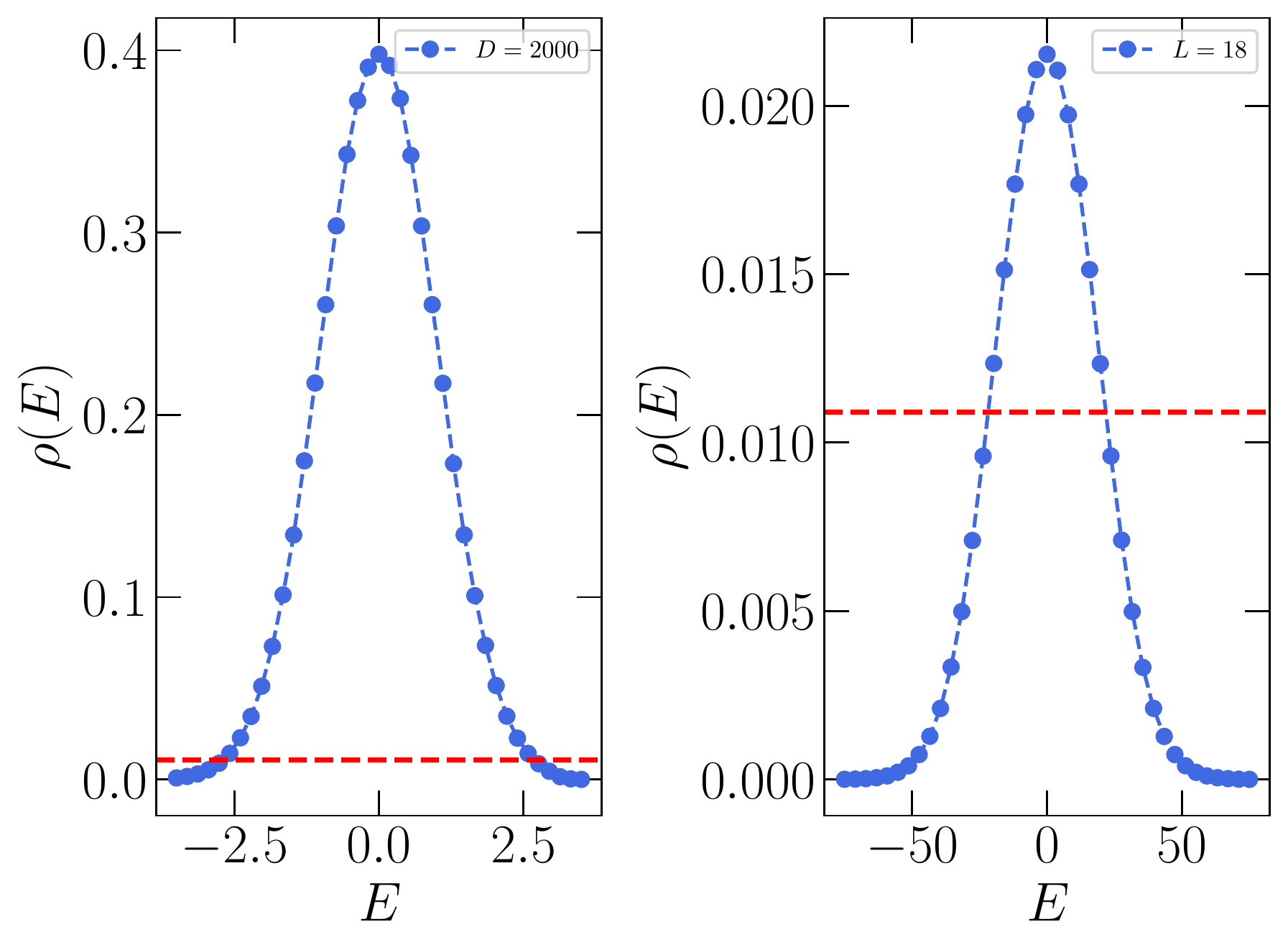}
 \vspace{-0.6cm}
 \caption{The normalized density of states of the combined ensemble ${\sf M}_i + \Lambda{\sf M_e}$ ($\Lambda = 1/D =  1/2000$) (left panel), and of the physical system $\hat H = \hat T + \hat V$, with $\xi = 2$ and $L=18$ (right panel). The critical density of states marking the regime below which the effective weight $\lambda$ of the ergodic matrix becomes negligible is marked by the horizontal dashed lines.}
 \label{fig:S3}
\end{figure}

\paragraph{The effective ergodic weight in different parts of the spectrum.---} Figure~\ref{fig:5} in the main text shows that for a mixed ensemble, the effective weight of the ergodic matrix on the eigenvalues of the mixed ensemble crucially depends on the density of the states of the latter. That is, for a denser region of the spectrum, the ergodicity sets in more strongly and is finite almost as any finite density is obtained for the eigenvalue spectrum. For the case of the physical system, where the dense GOE-matrix is substituted by a sparse non-random (but symmetric matrix), there is a critical density above which the ergodic aspects become relevant. We now analyze the matrix size dependence on these result in Fig.~\ref{fig:S2}. For the case of the pure mixed ensemble, Fig.~\ref{fig:S2}(a) shows that the linear relationship between the effective weight and the (normalized) density of states is valid for a wide range of matrix sizes $D$. Similarly, when using the physical model $\hat H = \hat T + \hat V$ [Fig.~\ref{fig:S2}(b)], the finite size effects do not qualitatively alter the main aspect expressed in the main text, in which a critical level density is necessary for the effective ergodic weight to become relevant. It is worth mentioning that the critical density is similar in all system sizes analyzed.

Finally, to see that this threshold density of eigenlevels is not vanishingly small, we report in Fig.~\ref{fig:S3} the normalized density of states of both cases, marking with a horizontal dashed line the maximum density where the effective weight $\lambda$ goes to zero. For the physical model, it is clear that a substantial part of the spectrum is thus not affected by the ergodic properties, and one can consequently identify a clear separation in behavior in different parts of it, which we interpret as the manifestation of a mobility edge.

\end{document}